\documentclass[aps, pre, twocolumn, floatfix]{revtex4-2}
\usepackage{graphicx, afterpage, amsmath, amssymb, color}
\usepackage{multirow, makecell}
\usepackage{hyperref}
\hypersetup{breaklinks={true}}

\renewcommand\dbltopfraction{1}
\renewcommand\textfraction{0}

\begin{document}
\title{$K$-selective percolation: A simple model leading to a rich repertoire of phase transitions}
\author{Jung-Ho Kim}
\author{K.-I. Goh}
\email{kgoh@korea.ac.kr}
\affiliation{Department of Physics, Korea University, Seoul 02841, Korea}
\date{\today}
\begin{abstract}
We propose the $K$-selective percolation process as a model for the iterative removals of nodes with the specific intermediate degree in complex networks.
In the model, a random node with degree $K$ is deactivated one by one until no more nodes with degree $K$ remain.
The non-monotonic response of the giant component size on various synthetic and real-world networks implies a conclusion that a network can be more robust against such selective attack by removing further edges.
In the theoretical perspective, the $K$-selective percolation process exhibits a rich repertoire of phase transitions, including double transitions of hybrid and continuous, as well as reentrant transitions.
Notably, we observe a tricritical-like point on Erd\H{o}s-R\'enyi networks.
We also examine a discontinuous transition with unusual order parameter fluctuation and distribution on simple cubic lattices, which does not appear in other percolation models with cascade processes.
Finally, we perform finite-size scaling analysis to obtain critical exponents on various transition points, including those exotic ones.
\end{abstract}
\maketitle

{\bf \noindent 
How robust is a system against attacks on nodes with a specific number of links?
To answer this deceptively simple question, we built a $K$-selective percolation model, which reveals surprisingly rich results.
From the practical network-scientific point of view, we found the possibility of an attack countermeasure which is to deactivate more edges.
This countermeasure works on various synthetic and real-world networks.
On the other hand, from the theoretical statistical physics perspective, a plethora of phase transitions appear, including a tricritical-like point and an exotic discontinuous phase transition.
Furthermore, we obtain a new set of critical exponents using finite-size scaling analysis.
}

\section{Introduction}
Since the 2000s, the network theory has become one of the most important theoretical toolboxes to understand the complex systems~\cite{2018NewmanNetworks}.
Percolation theory provides a theoretical foundation for how complex networks react to random failures and intentional attacks~\cite{2000ErrorAlbert}.
Knowing the response of the complex networks upon the attacks reveals the role of attacked nodes in the complex networks~\cite{2010CohenComplex, 2014AraujoRecent, 2015D'SouzaAnomalous, 2018LeeRecent, 2019D'SouzaExplosive, 2021LiPercolation}.
Most researches hitherto have been conducted either on the attacks to the high-degree nodes as in the earliest optimal percolation processes~\cite{2000ErrorAlbert, 2000CallawayNetwork, 2001CohenBreakdown, 2002HolmeAttack, 2010CohenComplex} or on the attacks to the low-degree nodes as in the $k$-core percolation process~\cite{1979ChalupaBootstrap, 2006DorogovtsevK, 2016LeeCritical}.
As yet, however, the robustness of the complex networks against the attacks on the specific intermediate-degree nodes still remains unaddressed.
One can easily think of various examples of attacks on the nodes with a specific intermediate degree.
One example is the collapse of the middle-class in modern socioeconomic networks~\cite{2009SchweitzerEconomic}.

\begin{figure}[b]
\centering
\vspace*{-0.4cm}
\includegraphics[width=0.5\textwidth]{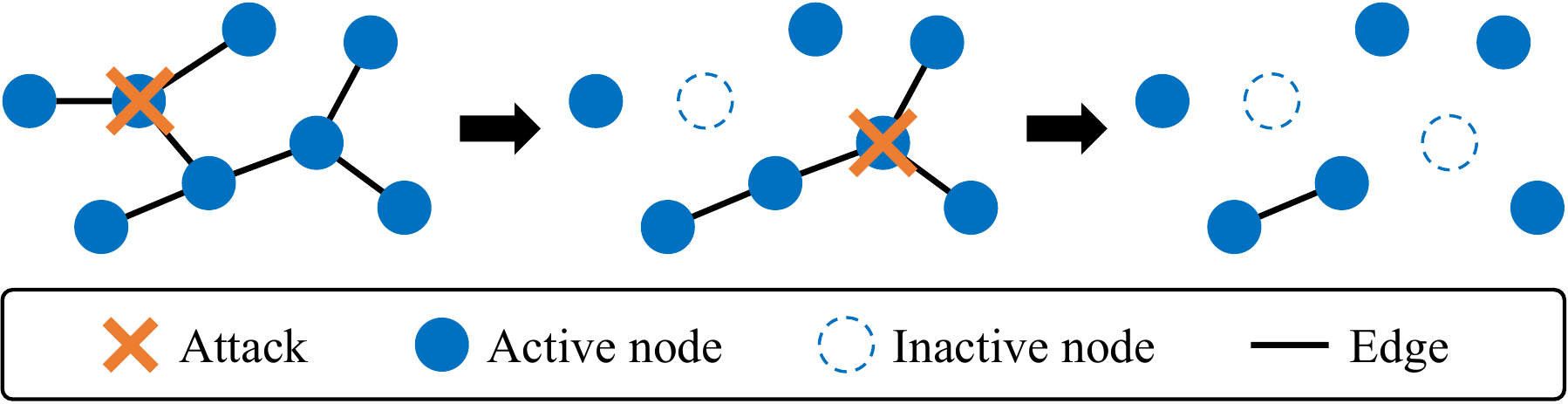}
\vspace*{-0.6cm}
\caption{A schematic illustration of the 3-selective percolation process on a simple network.}
\label{Fig:Model}
\end{figure}

To this end, we formulate a new percolation model, called $K$-selective percolation, to observe how the complex networks respond to the attacks on the specific degree nodes.
A schematic illustration of the 3-selective percolation process is shown in Fig.~\ref{Fig:Model}.
The $K$-selective percolation process proceeds with the following simple rules.
First, in a given initial network, all nodes are activated and the edges are activated independently with probability ($1-q$). 
We define the degree of a node as the number of active neighbor nodes connected by active edges.
Second, a random node with degree $K$ is chosen and deactivated.
Each time the node with degree $K$ is deactivated, its associated edges are also deactivated, and the degree of neighbors decreases by one.
Such $K$-selective node removal is repeated until there remain no more nodes with degree $K$ in the given network.

Let us make some remarks on the model.
i) This model is more `selective' than $k$-core percolation~\cite{1979ChalupaBootstrap, 2006DorogovtsevK, 2016LeeCritical} that makes every node with a degree less than $k$ deactivate successively.
ii) The outcome of the $K$-selective process is not unique but history-dependent.
However, the fluctuation of the giant component size caused by history-dependency tends to vanish in the thermodynamic limit.
iii) Parallel removal leads to a new model with different behavior.
iv) One can generalize the model into a limited-range $K$-selective percolation model without changing the main characteristics.

We use the probability $q$ that each edge of the given network is initially deactivated as the control parameter and the probability $G$ that a randomly chosen node belongs to the giant component as the order parameter, as in the usual percolation problem~\cite{1992StaufferIntroduction}.
We applied the $K$-selective percolation process to a wide range of substrates, including Erd\H{o}s-R\'enyi (ER) networks~\cite{1960ErdosOn}, three-dimensional simple cubic lattices, a polymer network model~\cite{2016KryvenRandom}, and the arXiv Condensed Matter collaboration network~\cite{2007LeskovecGraph}.

In most percolation models, the larger is $q$, the sparser the final network becomes.
Therefore the order parameter $G$ tends to decrease monotonically as a function of the control parameter $q$ and becomes zero at a unique percolation transition point.
The phase transition can be either continuous as in ordinary percolation~\cite{1992StaufferIntroduction} or hybrid as in $k$-core percolation~\cite{1979ChalupaBootstrap, 2006DorogovtsevK, 2016LeeCritical}.
Interestingly, it turns out that for $K$-selective percolation the behaviors of the order parameter $G$ are not that simple but rather more complicated.
One can typically observe upon increasing $q$ what we call the ``fragile valley", in which the system's vulnerability becomes locally maximal, followed by the ``resurgent hill", along which the system regains resilience.
These structures suggest that removing more edges could render a complex network more robust, contrary to common intuition~\cite{2004MotterCascade}.
One can also observe typically double phase transitions in this course.

We obtained the critical exponents of these phase transitions by finite-size scaling analysis.
We concluded that continuous phase transitions of the $K$-selective percolation process on both ER networks and the simple cubic lattices belong to the same universality class with ordinary percolation~\cite{1992StaufferIntroduction, 2013WangBond}.
For hybrid phase transition on ER networks, we obtained the critical exponents consistent with $k$-core percolation~\cite{2016LeeCritical} and cascading failure on multiplex networks~\cite{2016LeeHybrid}.
Besides, some unconventional transitions were observed as well.
For the 2-selective percolation process on ER networks, there appears a tricritical-like point, which has a unique critical exponent set.
For the 3-selective percolation process on simple cubic lattices, discontinuous phase transition with abnormal order parameter distribution and fluctuation is found, at which statistical distribution of the order parameter exhibits a one-sided power-law tail and the fluctuation of the order parameter diverges in the thermodynamic limit.

This paper is organized as follows.
First, we construct the numerical solution of the $K$-selective percolation process on random networks in Sec. \ref{Sec:Solution}.
Next, we demonstrate the main results of various synthetic and real-world networks in Sec. \ref{Sec:Results} and the finite-size scaling results of ER networks and simple cubic lattices in Sec. \ref{Sec:FSS}.
Finally, we make a conclusion in Sec. \ref{Sec:Conclusion}.

\section{Numerical solution}
\label{Sec:Solution}
We first present the numerical solution for the $K$-selective percolation process, applicable to random locally tree-like networks.
We derive a numerical solution for the evolution of the degree distribution of random networks during the $K$-selective percolation process using rate equations, which are similar to those of high-degree adaptive percolation~\cite{2020KimCritical}.
Here, the degree distribution $p_{k}$ is the probability that a randomly chosen node has degree $k$.

First, let us rescale the timescale of the process so that the rate at which the random nodes with degree $K$ deactivated is unity.
Then, $p_{K}$ and $p_{0}$ evolve in time as follows:
\begin{equation}
\frac{dp_{0}}{dt}=+1, ~~~~~~\frac{dp_{K}}{dt}=-1.
\label{Eq:dp0K}
\end{equation}
When a node with degree $K$ is deactivated, the degree of nodes linked to this deactivated node decreases by one.
The probability that a node connected to this deactivated node has degree $k$ is $kp_{k}/{\sum_{k'=0}^{\infty}{k'p_{k'}}}$.
This effect applies to every $k$, leading to
\begin{equation}
\frac{dp_{k}}{dt}=K\frac{(k+1)p_{k+1}}{\sum_{k'=0}^{\infty}{k'p_{k'}}}~ - ~K\frac{kp_{k}}{\sum_{k'=0}^{\infty}{k'p_{k'}}}.
\label{Eq:dpk}
\end{equation}
On the right-hand side, the first, positive term represents the nodes with degree $(k+1)$ linked to the deactivated node; and the second, negative term represents the nodes with degree $k$ linked to the deactivated node.
It is worth noticing that the denominator ${\sum_{k'=0}^{\infty}{k'p_{k'}}}$ is not constant in time but rather a decreasing function of time, as we count the degree of a node by the number of active neighbor nodes.

The complete set of rate equations can be obtained by combining Eq.~(\ref{Eq:dp0K}) and Eq.~(\ref{Eq:dpk}):
\begin{equation}
\begin{aligned}
\frac{dp_{0}}{dt}=&~K\frac{p_{1}}{\sum_{k'=0}^{\infty}{k'p_{k'}}}~ + ~1,\\
\frac{dp_{1}}{dt}=&~K\frac{(2p_{2} - p_{1})}{\sum_{k'=0}^{\infty}{k'p_{k'}}},\\
\qquad\qquad\vdots \\
\frac{dp_{K}}{dt}=&~K\frac{[(K+1)p_{K+1} - Kp_{K}]}{\sum_{k'=0}^{\infty}{k'p_{k'}}}~ - ~1,\\
\qquad\qquad\vdots \\
\frac{dp_{k}}{dt}=&~K\frac{[(k+1)p_{k+1} - kp_{k}]}{\sum_{k'=0}^{\infty}{k'p_{k'}}},\\
\qquad\qquad\vdots \\
\end{aligned}
\end{equation}
These rate equations should be iterated until $p_{K}=0$, which means that there remain no more nodes with degree $K$ in the given network.

\begin{figure}[t]
\centering
\includegraphics[width=0.5\textwidth]{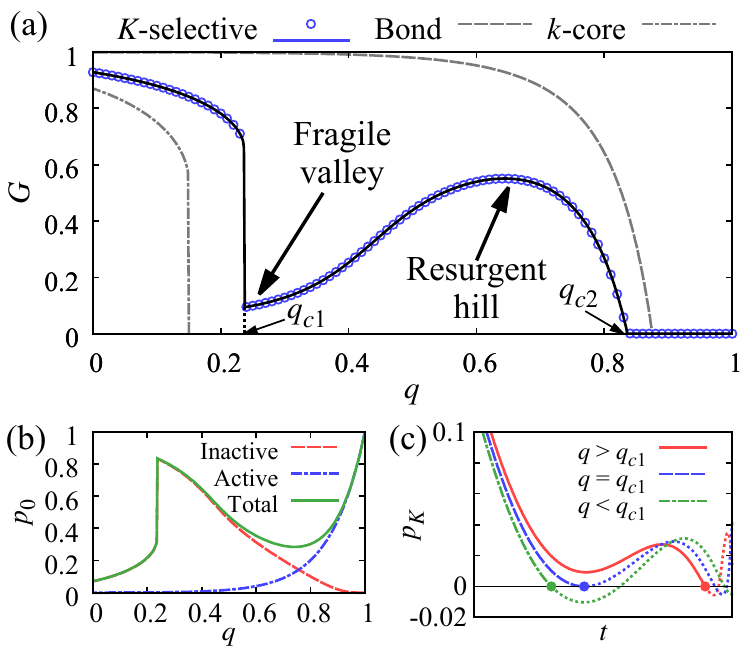}
\vspace*{-0.6cm}
\caption{Typical $K$-selective percolation results with $K=4$ on ER networks with mean degree $z=8$.
(a) Plots of the order parameter $G$ as a function of $q$ for the bond, $k$-core, and $K$-selective percolation processes with $k=5$ and $K=4$, respectively.
The lines are the analytic and numerical solutions, and the points are the Monte Carlo simulation results with the system size $N=10^{7}$.
(b) Plots of $p_{0}$ after the $K$-selective percolation process, as a function of $q$.
$p_{0}$ consists of two kinds of nodes, the inactive nodes and the active nodes with degree zero.
(c) Plots of $p_{K}$ during the $K$-selective percolation process near the $q_{c1}$.
Time $t$ has an arbitrary unit.
The percolation process ends at the round markers, and dotted lines are physically unreachable.}
\vspace*{-0.2cm}
\label{Fig:Result01}
\end{figure}

Various quantities of interest, including the order parameter $G$, can be calculated by formal generating function method with degree distribution $p_{k}$, under the assumption that after the $K$-selective percolation process, the remaining network still can be characterized solely by $p_{k}$, like tree-like random networks~\cite{2000CallawayNetwork, 2001NewmanRandom}.
The agreement of the Monte Carlo simulation results and this numerical solution as shown in Fig.~\ref{Fig:Result01}(a) supports the validity of this assumption.

\begin{figure}[t]
\includegraphics[width=0.5\textwidth]{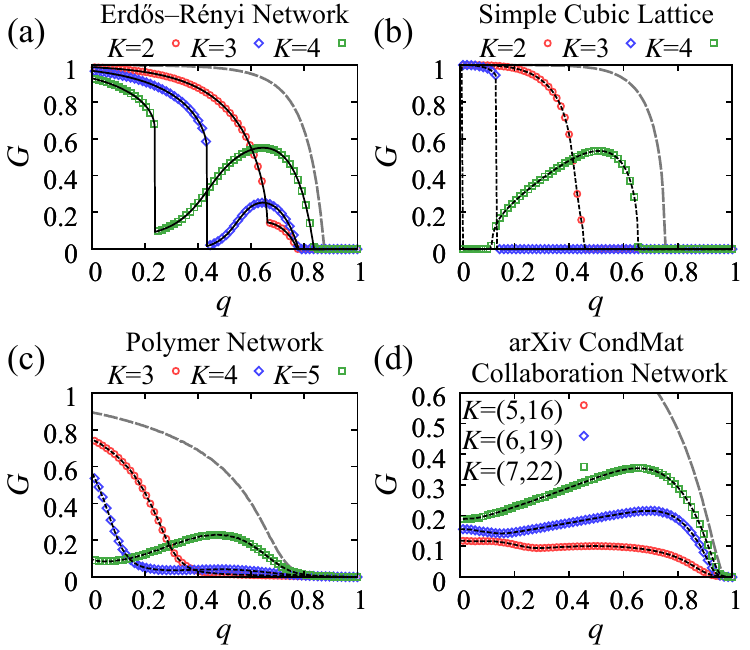}
\vspace*{-0.6cm}
\caption{Plots of the order parameter $G$ as a function of $q$ for the $K$-selective percolation process (a) on ER networks with mean degree $z=8$ and the system size $N=10^{7}$, (b) on three-dimensional simple cubic lattices with $N=464^{3}$, (c) on the polymer networks~\cite{2016KryvenRandom} with $z\simeq6.28$ and $N=5000$, and (d) on the arXiv Condensed Matter collaboration network~\cite{2007LeskovecGraph} with $z\simeq8.08$ and $N=23133$.
Points are the Monte Carlo simulation results.
Solid lines in (a) are the results of the numerical solution, and dotted lines on (b--d) are mere guidelines.
Gray dashed lines in (a--d) are bond percolation results.}
\vspace*{-0.2cm}
\label{Fig:Result02}
\end{figure}

\section{Results}
\label{Sec:Results}
Typical results of the $K$-selective percolation are shown in Fig.~\ref{Fig:Result01}.
These results were obtained for $K$-selective percolation with $K=4$ on ER networks with mean degree $z=8$.
There are two phase transition points $q_{c1}=0.236\,262\,511$ and $q_{c2}=0.836\,137\,053$.
The first noteworthy feature is the fragile valley and resurgent hill structure indicated in Fig.~\ref{Fig:Result01}(a).
The fragile valley is where the system becomes suddenly vulnerable to attacks on the nodes with degree $K$; and the resurgent hill is where the system regains resilience to attacks on the nodes with degree $K$.
An intriguing question about the network's robustness arises here.
If a complex network is in the fragile valley, what can we do to make it more robust?
The first solution is to restore more edges (decreasing $q$), which is mathematically trivial but often expensive and impractical.
The other way is to remove some edges further (increasing $q$).
This solution places the system upward to the resurgent hill. 
Even though the number of edges is decreased, and thus the network becomes sparser, the network nevertheless can withstand better against the attacks~\cite{2004MotterCascade}.

\begin{table*}[t]
\centering
\begin{tabular}{>{\centering}m{0.12\textwidth} >{\centering}m{0.20\textwidth} >{\centering}m{0.06\textwidth} >{\centering}m{0.13\textwidth} >{\centering}m{0.15\textwidth} >{\centering}m{0.08\textwidth} >{\centering}m{0.08\textwidth} >{\centering}m{0.08\textwidth}}
\hline\hline
Network & Percolation Model & $K$ & $z_{c}$ & Transition Type & $\beta$ & $\gamma$ & $\bar{\nu}$ \tabularnewline
\hline
\multirow{6}{*}{\shortstack{Erd\H{o}s-R\'enyi\\network}} & \multirow{3}*{$K$-selective percolation} & 3 & 1.770\,502\,87 & Continuous & 0.99(5) & 0.99(5) & 2.97(14) \tabularnewline
& & 3 & 4.535\,985\,41 & Hybrid & 0.49(2) & 0.96(4) & 2.03(7) \tabularnewline
& & 2 & 2.718\,28 & Tricritical-like & 0.50(1) & 1.78(4) & 2.86(6) \tabularnewline
\cline{2-8}
& \multicolumn{3}{c}{Ordinary percolation~\cite{1992StaufferIntroduction}} & Continuous & 1 & 1 & 3 \tabularnewline
& \multicolumn{3}{c}{$k$-core percolation~\cite{2016LeeCritical}} & Hybrid & 0.50(1)  & 0.97(1) & 2.06(5) \tabularnewline
& \multicolumn{3}{c}{Cascading failure on multiplex networks~\cite{2016LeeHybrid}} & Hybrid & 0.50(1) & 1.05(5) & 2.10(2) \tabularnewline
\hline\hline
\end{tabular}
\caption{Critical points and critical exponents for $K$-selective, ordinary, and $k$-core percolation on ER networks and cascading failure model on multiplex ER networks.}
\label{Table:ERExponents}
\end{table*}

The origin of the fragile valley and the resurgent hill structure is suggested in Fig.~\ref{Fig:Result01}(b).
To this end, we examine the behavior of $p_{0}$ which is contributed by two kinds of nodes: Firstly, the directly-inactive nodes (red dashed line) that are deactivated directly by the $K$-selective percolation attacks, and secondly the active nodes (blue dashed-dot line) with degree zero.
The inactive nodes fraction peaks at the fragile valley, because near $q_{c1}$ there exist many nodes with degree $K$ or to be $K$ during the $K$-selective percolation process.
On the other hand, the fraction of active nodes with degree zero peaks at $q=1$, because there are no active edges.
High $p_{0}$ makes the network sparser, and the giant component size shrink.
Therefore these two $p_{0}$ peaks suggest the fragile valley and the resurgent hill structure.

The underlying mechanism for the discontinuity at the first phase transition point $q_{c1}$ is proposed in Fig.~\ref{Fig:Result01}(c).
Near $q_{c1}$, the plots of $p_{K}$ as a function of time are S-shaped; reminiscent of the $p$-$V$ curve of the van der Waals gas.
When $q<q_{c1}$ (green dashed-dot line), the $p_{K}$ curve touches zero before the local minimum of the S-shape.
As soon as $p_{K}$ touches zero, there remain no more nodes with degree $K$ and the $K$-selective percolation process is over.
Therefore, the dotted lines after $p_{K}$ touches zero are unphysical.
At the critical point $q_{c1}$ (blue dashed line), the $p_{K}$ curve touches zero exactly at the local minimum.
Upon slightly increasing $q$ beyond the critical point (red solid line), the $K$-selective percolation process can proceed beyond the first local minimum, resulting in many more nodes being deactivated.
This induces the discontinuity of the order parameter.

We performed the $K$-selective percolation process on ER networks, three-dimensional simple cubic lattices with periodic boundary conditions, a polymer network model~\cite{2016KryvenRandom}, and the arXiv Condensed Matter collaboration network~\cite{2007LeskovecGraph}.
As summarized in Fig.~\ref{Fig:Result02}, the phenomenology from all these diverse networks is qualitatively similar.
Double phase transitions appear on ER networks [Fig.~\ref{Fig:Result02}(a)], and for the 2-selective percolation, there exists a tricritical-like point.
For the simple cubic lattices [Fig.~\ref{Fig:Result02}(b)], discontinuous phase transition appears for the 3-selective percolation process, and reentrant phase transition appears for the 4-selective percolation process.
One can also observe the valley-hill structure on the polymer networks [Fig.~\ref{Fig:Result02}(c)] and the arXiv Condensed Matter collaboration network [Fig.~\ref{Fig:Result02}(d)].
For the polymer networks, we used the model proposed by Kryven \textit{et al.}~\cite{2016KryvenRandom}.
The degree distribution of the polymer network comes from the real chemical data~\cite{2016KryvenRandom}.
Detailed model descriptions and parameters we used are provided in the Appendix \ref{App:Polymer}.
We applied the limited-range $K$ on the collaboration network~\cite{2007LeskovecGraph}, which has power-law degree distribution.
For instance, for $K=(5,16)$ in Fig.~\ref{Fig:Result02}(d), we select and deactivate a random node with degrees $5 \leq k \leq 16$ iteratively.

\section{Finite-size scaling analysis}
\label{Sec:FSS}
We examine the phase transitions on ER networks (Sec. \ref{Sec:FSSER}.) and three-dimensional simple cubic lattices with periodic boundary conditions (Sec. \ref{Sec:FSSCubic}.) for the $K$-selective percolation process in more detail.
For continuous phase transition to eventual disappearance of the giant component in both ER and simple cubic lattice, we obtained the critical exponents same as ordinary percolation~\cite{1992StaufferIntroduction, 2013WangBond}; for hybrid phase transition observed in ER, we obtained critical exponents consistent with those of $k$-core percolation~\cite{2016LeeCritical} and cascading failure on multiplex networks~\cite{2016LeeHybrid}.
On the contrary, we found a unique set of critical exponents for the tricritical-like point of the ER network (Sec. \ref{Sec:Tricritical-like}.), and discontinuous phase transition of simple cubic lattices has abnormal order parameter distribution and fluctuation (Sec. \ref{Sec:Discontinuous}.).

\subsection{ER networks}
\label{Sec:FSSER}
In this subsection, we use the mean degree $z$ of diluted ER network as the control parameter for convenience.
The control parameter $q$ and $z$ have the relation; $z=z_{0} \times (1-q)$ where $z_{0}=8$ here.
In the vicinity of the critical point, the order parameter $G$ and the order parameter fluctuation $\chi \equiv N(\langle G^{2} \rangle - \langle G \rangle ^{2})$ are known to exhibit power laws with critical exponents $\beta$ and $\gamma$ respectively,
\begin{eqnarray}
\label{Eq:ERBeta}
G(z)-G(z^{+}_{c}) \propto & (z_{c}-z)^{\beta} ~~~~& \textrm{with}~z\rightarrow z_{c}^{+}~,\\
\label{Eq:ERGamma}
\chi(z) \propto & (z_{c}-z)^{-\gamma} ~~~~& \textrm{with}~z\rightarrow z_{c}^{+}~,
\end{eqnarray}
where $G(z^{+}_{c})$ is the order parameter just above the phase transition.
$G(z^{+}_{c})$ remains nonzero for hybrid phase transitions and approaches zero for continuous phase transitions.

We use finite-size scaling ansatz and Monte Carlo simulations to obtain these critical exponents.
According to the finite-size scaling theory, percolation quantities like $G$ and $\chi$ have the scaling form near the critical point with the system size $N$ as
\begin{eqnarray}
\label{Eq:ERScalingAnsatz1}
G(N, z) - G(z^{+}_{c}) = & N^{-\beta/\bar{\nu}} \widetilde{G}[(z_{c}-z)N^{1/\bar{\nu}}]~,\\
\label{Eq:ERScalingAnsatz2}
\chi(N, z) = & N^{\gamma/\bar{\nu}} \widetilde{\chi}[(z_{c}-z)N^{1/\bar{\nu}}]~,
\end{eqnarray}
where $\widetilde{G}$ and $\widetilde{\chi}$ are the scaling functions and $\bar\nu$ is the finite-size scaling exponent.
For continuous phase transitions, $\nu$ is the correlation length critical exponent with $\xi(z) \propto (z_{c}-z)^{-\nu}$ in the vicinity of the critical point, and $\bar\nu=d\nu$ with $d$ being an effective dimension~\cite{2007HongFinite}.
The obtained critical points and critical exponents are summarized in Table \ref{Table:ERExponents} with those of the other percolation processes.

\begin{figure*}[t]
\centering
\includegraphics[width=0.95\textwidth]{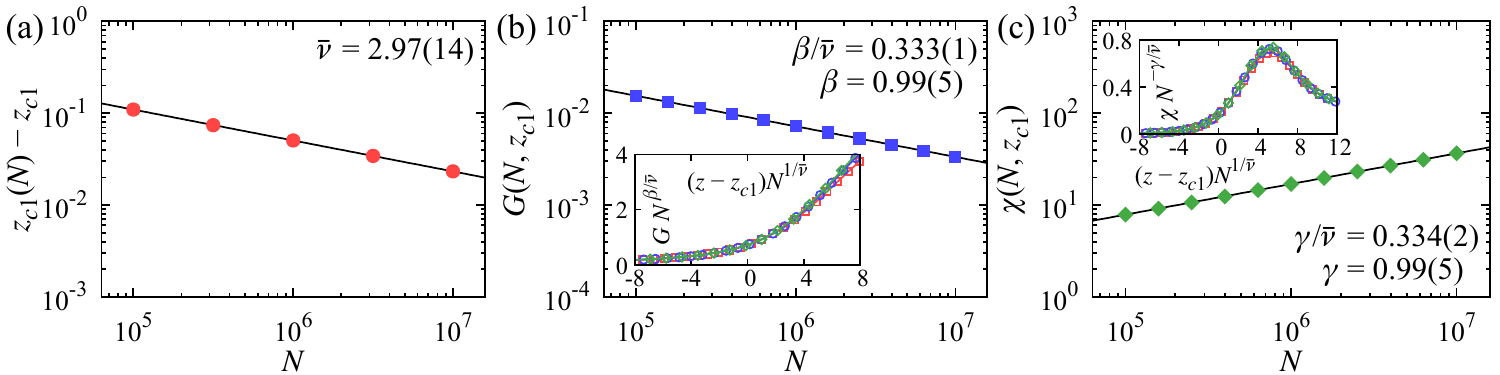}
\vspace*{-0.4cm}
\caption{Finite-size scaling analysis results for 3-selective percolation on ER networks at a continuous phase transition point $z_{c1}=1.770\,502\,87$.
(a) represents the relation of Eq.~(\ref{Eq:FSSER31n}), (b) represents the relation of Eq.~(\ref{Eq:FSSER31bn}), and (c) represents the relation of Eq.~(\ref{Eq:FSSER31gn}).
Points represent Monte Carlo simulation results, and black solid lines are the fitting lines.
(Insets) Plots of finite-size-scaled data collapse curve from Eqs.~(\ref{Eq:ERScalingAnsatz1}--\ref{Eq:ERScalingAnsatz2}), using the obtained critical exponents.
Monte Carlo simulation results with system size $N=10^{5}$ (square), $10^{6}$ (circle), and $10^{7}$ (diamond) are used for the inset plots.}
\vspace*{-0.2cm}
\label{Fig:FSSER01}
\end{figure*}

\subsubsection{Continuous phase transition}
For 3-selective percolation on ER networks, there is a continuous phase transition point at $z_{c1}=1.770\,502\,87$, which was obtained from the numerical solution.
We obtained critical exponents from the following relations, which come from the Eqs.~(\ref{Eq:ERScalingAnsatz1}--\ref{Eq:ERScalingAnsatz2}):
\begin{eqnarray}
\label{Eq:FSSER31n}
z_{c1}(N)-z_{c1} &\propto& N^{-1 / \bar\nu},\\
\label{Eq:FSSER31bn}
G(N, z_{c1}) &\propto& N^{-\beta / \bar\nu},\\
\label{Eq:FSSER31gn}
\chi(N, z_{c1}) &\propto& N^{\gamma / \bar\nu},
\end{eqnarray}
where $z_{c1}(N)$ is defined as a point with the maximum value of $\chi$ on ER networks with system size $N$.
As shown in Fig.~\ref{Fig:FSSER01}, we obtained $\beta=0.99(5)$, $\gamma=0.99(5)$ and $\bar\nu=2.97(14)$, and the rescaled data using these exponents are collapsed well onto a single line.
The obtained exponents are the same as those of ordinary percolation within the margin of errors (See Table \ref{Table:ERExponents}).\

\subsubsection{Hybrid phase transition}
For 3-selective percolation on ER networks, there is a hybrid phase transition point at $z_{c2}=4.535\,985\,41$ with $G(z^{-}_{c2})=0.016\,409\,2$ and $G(z^{+}_{c2})=0.537\,21$, which were obtained from the numerical solution.
First, to find $\bar\nu$, we rescale the control parameter $r=z/(2 \times z_{c2})$. 
We obtained fixed point $r^{*}_{c2}(N)$ for large cell renormalization group transformation~\cite{1980ReynoldsLarge}, satisfying $r^{*}_{c2}(N)=\Pi(N,r^{*}_{c2})$.
Generally, $\Pi(N, r^{*}_{c2})$ is the probability that there exists the percolating cluster in the lattice percolation problem.
However, $\Pi(N, r^{*}_{c2})$ could be obtained for the hybrid phase transition on the networks by the probability that the network is in the upper branch of the order parameter curve.
As shown in Fig.~\ref{Fig:FSSER02}(a), one can easily distinguish whether the network is in the upper branch (right hump) or in the lower branch (left hump), even if the system is finite.

According to large cell renormalization group theory~\cite{1980ReynoldsLarge}, $\bar\nu$ could be obtained from following equation:
\begin{eqnarray}
\label{Eq:FSSER32n}
\frac{d\Pi(N,r)}{dr}\bigg\rvert_{r=r^{*}_{c2}(N)} &\propto N^{1 / \bar\nu}.
\end{eqnarray}
As shown in Fig.~\ref{Fig:FSSER02}(b), we obtained $\bar\nu=2.03(7)$ and the rescaled data using these exponents are collapsed well onto a single line.
In general, $\bar\nu$ is the same as $\bar\nu^{\prime}$ from $r_{c2}-r^{*}_{c2}(N) \propto N^{-1 / \bar\nu^{\prime}}$.
However, as shown in Figs.~\ref{Fig:FSSER02}(c), $\bar\nu$ and $\bar\nu^{\prime}$ are different from each other and the rescaled data using  $\bar\nu^{\prime}$ are not collapsed well.
This phenomenon has also been reported for hybrid phase transition on cascading failure on multiplex networks~\cite{2016LeeHybrid}, but not for hybrid phase transition on $k$-core percolation~\cite{2016LeeCritical}.

Next, we obtained other critical exponents from the following relations, which come from the Eqs.~(\ref{Eq:ERScalingAnsatz1}--\ref{Eq:ERScalingAnsatz2}):
\begin{eqnarray}
\label{Eq:FSSER32bn}
G(N, z_{c2})-G(z^{+}_{c2}) &\propto& N^{-\beta / \bar\nu},\\
\label{Eq:FSSER32gn}
\chi(N, z_{c2}) &\propto& N^{\gamma / \bar\nu}.
\end{eqnarray}
As shown in Fig.~\ref{Fig:FSSER02}(d--e), we obtained $\beta=0.49(2)$ and $\gamma=0.96(4)$ and the rescaled data using these exponents are collapsed well onto a single line.
Note that data for Figs.~\ref{Fig:FSSER02}(d--e), we only used Monte Carlo ensembles that belong to the upper branch of the order parameter curve.
The critical exponent $\beta$ obtained from the relation of Eq.~(\ref{Eq:ERBeta}) with the numerical solution supports obtained $\beta$ from finite-size scaling analysis.
These exponents are consistent with those of $k$-core percolation~\cite{2016LeeCritical} and cascading failure on multiplex networks~\cite{2016LeeHybrid} (See Table \ref{Table:ERExponents}).

\afterpage{
\onecolumngrid

\renewcommand\dbltopfraction{1}
\renewcommand\textfraction{0}
\begin{figure}[t]
\centering
\includegraphics[width=0.95\textwidth]{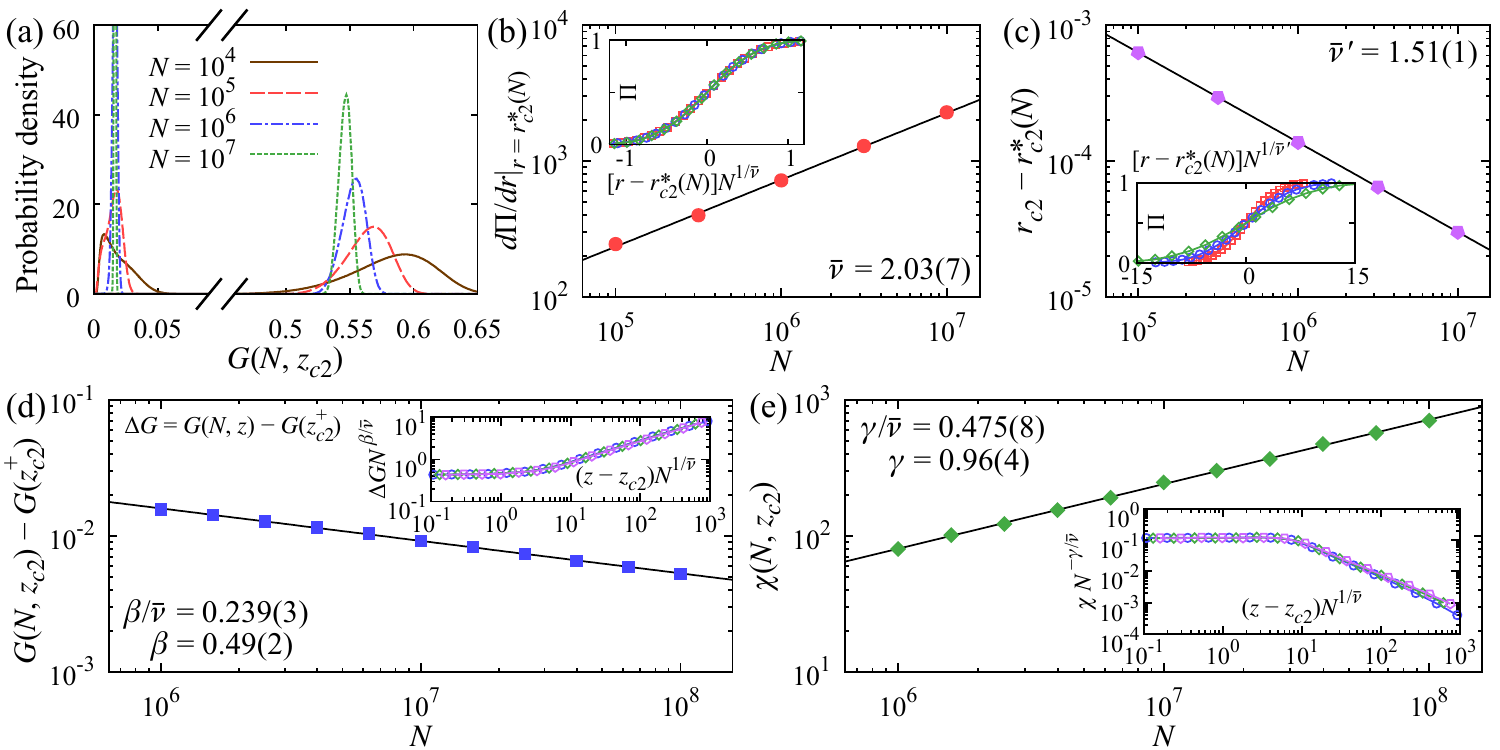}
\vspace*{-0.4cm}
\caption{Finite-size scaling analysis results for 3-selective percolation on ER networks at a hybrid phase transition point $z_{c2}=4.535\,985\,41$ with $G(z^{-}_{c2})=0.016\,409\,2$ and $G(z^{+}_{c2})=0.537\,21$.
(a) Plots of the probability density of the order parameter $G$ from the Monte Carlo simulation at the critical point.
(b) represents the relation of Eq.~(\ref{Eq:FSSER32n}), (c) represents the relation of $r_{c2}-r^{*}_{c2}(N) \propto N^{-1 / \bar\nu^{\prime}}$, (d) represents the relation of Eq.~(\ref{Eq:FSSER32bn}), and (e) represents the relation of Eq.~(\ref{Eq:FSSER32gn}).
For (d--e) we only used Monte Carlo ensembles that belong to the right hump of (a).
(b--e) Points represent Monte Carlo simulation results, and black solid lines are the fitting lines.
(Insets) Plots of finite-size-scaled data collapse curve from $\Pi=\widetilde{\Pi}[(r-r^{*}_{c2}(N))N^{1/\bar\nu}]$ and Eqs.~(\ref{Eq:ERScalingAnsatz1}--\ref{Eq:ERScalingAnsatz2}), using the obtained critical exponents.
Note that data collapse is only achieved by using $r^{*}_{c2}(N)$ instead of $r_{c2}$ and $\bar\nu$ instead of $\bar\nu^{*}$.
Monte Carlo simulation results with system size $N=10^{5}$ (square), $10^{6}$ (circle), $10^{7}$ (diamond) and $10^{8}$ (pentagon) are used for the inset plots.}
\label{Fig:FSSER02}
\centering
\includegraphics[width=0.95\textwidth]{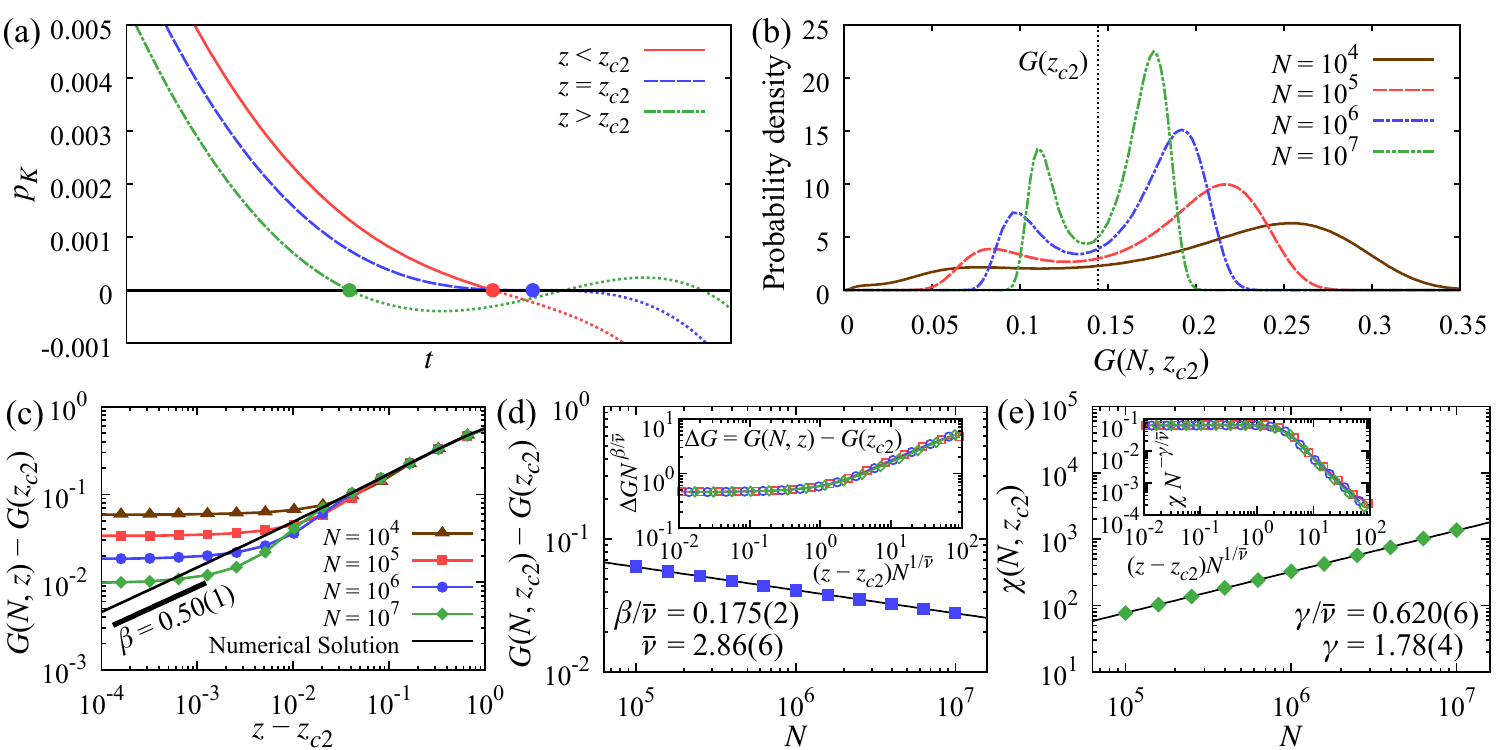}
\vspace*{-0.4cm}
\caption{Finite-size scaling analysis results for 2-selective percolation on ER networks at a tricritical-like point $z_{c2}=2.718\,28$ with $G(z_{c2})=0.144\,61$.
(a) Plots of $p_{K}$ during the $K$-selective percolation process near the critical point.
Time $t$ has an arbitrary unit.
The percolation processes end at the round markers and dotted lines are physically unreachable.
(b) Plots of the probability density of the order parameter $G$ from the Monte Carlo simulation at the tricritical-like point.
(c) represents the relation of Eq.~(\ref{Eq:ERBeta}).
Point lines are the Monte Carlo simulation results, and the solid line is the numerical solution.
(d) represents the relation of Eq.~(\ref{Eq:FSSER22bn}) and (e) represents the relation of Eq.~(\ref{Eq:FSSER22gn}).
(d--e) We only used Monte Carlo ensembles that belong to the upper branch.
Points represent Monte Carlo simulation results, and black solid lines are the fitting lines.
(Insets) Plots of finite-size-scaled data collapse curve from Eqs.~(\ref{Eq:ERScalingAnsatz1}--\ref{Eq:ERScalingAnsatz2}), using the obtained critical exponents.
Monte Carlo simulation results with system size $N=10^{5}$ (square), $10^{6}$ (circle) and $10^{7}$ (diamond) are used for the inset plots.}
\label{Fig:FSSER03}
\end{figure}
\cleardoublepage
\twocolumngrid}

\afterpage{
\onecolumngrid

\begin{table}[t]
\centering
\begin{tabular}{>{\centering}m{0.12\textwidth} >{\centering}m{0.20\textwidth} >{\centering}m{0.06\textwidth} >{\centering}m{0.13\textwidth} >{\centering}m{0.15\textwidth} >{\centering}m{0.08\textwidth} >{\centering}m{0.08\textwidth} >{\centering}m{0.08\textwidth}}
\hline\hline
Lattice & Percolation Model & $K$ & $p_{c}$ & Transition Type & $\beta$ & $\gamma$ & $\bar{\nu}$ \tabularnewline
\hline
\multirow{3}{*}{\shortstack{Simple cubic\\lattice}} & \multirow{2}*{$K$-selective percolation} & 4 & 0.346\,525(10) & Continuous & 0.418(2) & 1.80(1) & 2.64(1) \tabularnewline
& & 3 & 0.868\,32(3) & Discontinuous & - & 1.14(5) & 4.14(7) \tabularnewline
\cline{2-8}
& \multicolumn{3}{c}{Ordinary percolation~\cite{2013WangBond}} & Continuous & 0.4181(6) & 1.793(2) & 2.629(3) \tabularnewline
\hline\hline
\end{tabular}
\vspace*{-0.2cm}
\caption{Critical points and critical exponents for $K$-selective and ordinary percolation on three-dimensional simple cubic lattices with periodic boundary conditions.}
\label{Table:CubicExponents}
\end{table}
\twocolumngrid}

\subsubsection{Tricritical-like point}
\label{Sec:Tricritical-like}
For 2-selective percolation on ER networks, the intermediate transition becomes, instead of a conventional hybrid-type, a tricritical-like point at $z_{c2}=2.718\,28$ with $G(z_{c2}) = 0.144\,61$ [Fig~.\ref{Fig:Result02}(a)], which was obtained from the numerical solution.
We concluded this point is a tricritical-like point on the basis of the behavior of $p_{2}$ during the $2$-selective process [Fig.~\ref{Fig:FSSER03}(a)] and the order parameter distribution [Fig.~\ref{Fig:FSSER03}(b)] at this point.
When $z>z_{c2}$ (green dashed-dot line), $p_{2}$ displays S-shape like Fig.~\ref{Fig:Result01}(c), and at $z_{c2}$ (blue dashed line) the local minimum and the local maximum merge into one point.
Also, the order parameter distribution at $z_{c2}$ has two peaks, which get closer as the system size increases.
These two results suggest that $z_{c2}$ is a tricritical-like point.
At this tricritical-like point, the two peaks cannot be separated exactly.
Therefore, we assume that the ensembles with $G > G(z_{c2})$ [$G<G(z_{c2})$] belong to the upper (lower) branch at $z_{c2}$.
First, we obtained $\beta=0.50(1)$ from the relation of Eq.~(\ref{Eq:ERBeta}) with the numerical solution [Fig.~\ref{Fig:FSSER03}(c)].
Then, we obtained other critical exponents from the following relations, which come from the Eqs.~(\ref{Eq:ERScalingAnsatz1}--\ref{Eq:ERScalingAnsatz2}):
\begin{eqnarray}
\label{Eq:FSSER22bn}
G(N, z_{c2})-G(z_{c2}) &\propto& N^{-\beta / \bar\nu},\\
\label{Eq:FSSER22gn}
\chi(N, z_{c2}) &\propto& N^{\gamma / \bar\nu}.
\end{eqnarray}
As shown in Fig.~\ref{Fig:FSSER03}(d--e), we obtained $\bar\nu=2.86(6)$ and $\gamma=1.78(4)$ and the rescaled data using these exponents are collapsed well onto a single line.
As in hybrid phase transition, we only used Monte Carlo ensembles that belong to the upper branch.
These exponents are apparently different from any of hybrid and continuous phase transitions.

\subsection{Simple cubic lattices}
\label{Sec:FSSCubic}
In this subsection, we use the probability $p$ that each edge of the given lattice is initially active as the control parameter for convenience.
The control parameter $q$ and $p$ have the relation; $p=1-q$.
We used periodic boundary conditions for three-dimensional simple cubic lattices.
We defined the percolating cluster as a cluster that contains more than one site of every two-dimensional layer for a fixed direction.

The corresponding formulas for Eqs.~(\ref{Eq:ERBeta}--\ref{Eq:ERGamma}) and Eqs.~(\ref{Eq:ERScalingAnsatz1}--\ref{Eq:ERScalingAnsatz2}) are as follows:
\begin{eqnarray}
\label{Eq:CubicBeta}
G(p)-G(p^{+}_{c}) \propto & (p_{c}-p)^{\beta} ~~~~& \textrm{with}~p\rightarrow p_{c}^{+}~,\\
\label{Eq:CubicGamma}
\chi(p) \propto & (p_{c}-p)^{-\gamma} ~~~~& \textrm{with}~p\rightarrow p_{c}^{+}~,
\end{eqnarray}
\begin{eqnarray}
\label{Eq:CubicScalingAnsatz1}
G(N, p) - G(p^{+}_{c}) = & N^{-\beta/\bar{\nu}} \widetilde{G}[(p_{c}-p)N^{1/\bar{\nu}}]~,\\
\label{Eq:CubicScalingAnsatz2}
\chi(N, p) = & N^{\gamma/\bar{\nu}} \widetilde{\chi}[(p_{c}-p)N^{1/\bar{\nu}}]~.
\end{eqnarray}

The obtained critical points and critical exponents are summarized in Table \ref{Table:CubicExponents} with those of the ordinary percolation process.

\subsubsection{Continuous phase transition}
For 4-selective percolation on the simple cubic lattices, there is a continuous phase transition point at $p_{c1}=0.346\,525(10)$, which obtained from the crossing point of the percolation probability $\Pi$ for various system sizes [Fig.~\ref{Fig:FSSCubic01}(a)].
We obtained critical exponents from the following relations, which come from the Eqs.~(\ref{Eq:CubicScalingAnsatz1}--\ref{Eq:CubicScalingAnsatz2}):
\begin{eqnarray}
\label{Eq:FSSCubic41n}
p_{c1}-p^{*}(N) &\propto& N^{-1 / \bar\nu},\\
\label{Eq:FSSCubic41bn}
G(N, p_{c1}) &\propto& N^{-\beta / \bar\nu},\\
\label{Eq:FSSCubic41gn}
\chi(N, p_{c1}) &\propto& N^{\gamma / \bar\nu},
\end{eqnarray}
where $p^{*}(N)$ is the fixed point satisfying $p^{*}(N)=\Pi(N,p^{*})$.
As shown in Fig.~\ref{Fig:FSSCubic01}(b--d), we obtained $\beta=0.418(2)$, $\gamma=1.80(1)$ and $\bar\nu=2.64(1)$, and the rescaled data using these exponents are collapsed well onto a single line.
The obtained exponents are the same as those of ordinary percolation within the margin of errors (See Table \ref{Table:CubicExponents}).

\afterpage{
\onecolumngrid

\begin{figure}[t]
\centering
\includegraphics[width=0.95\textwidth]{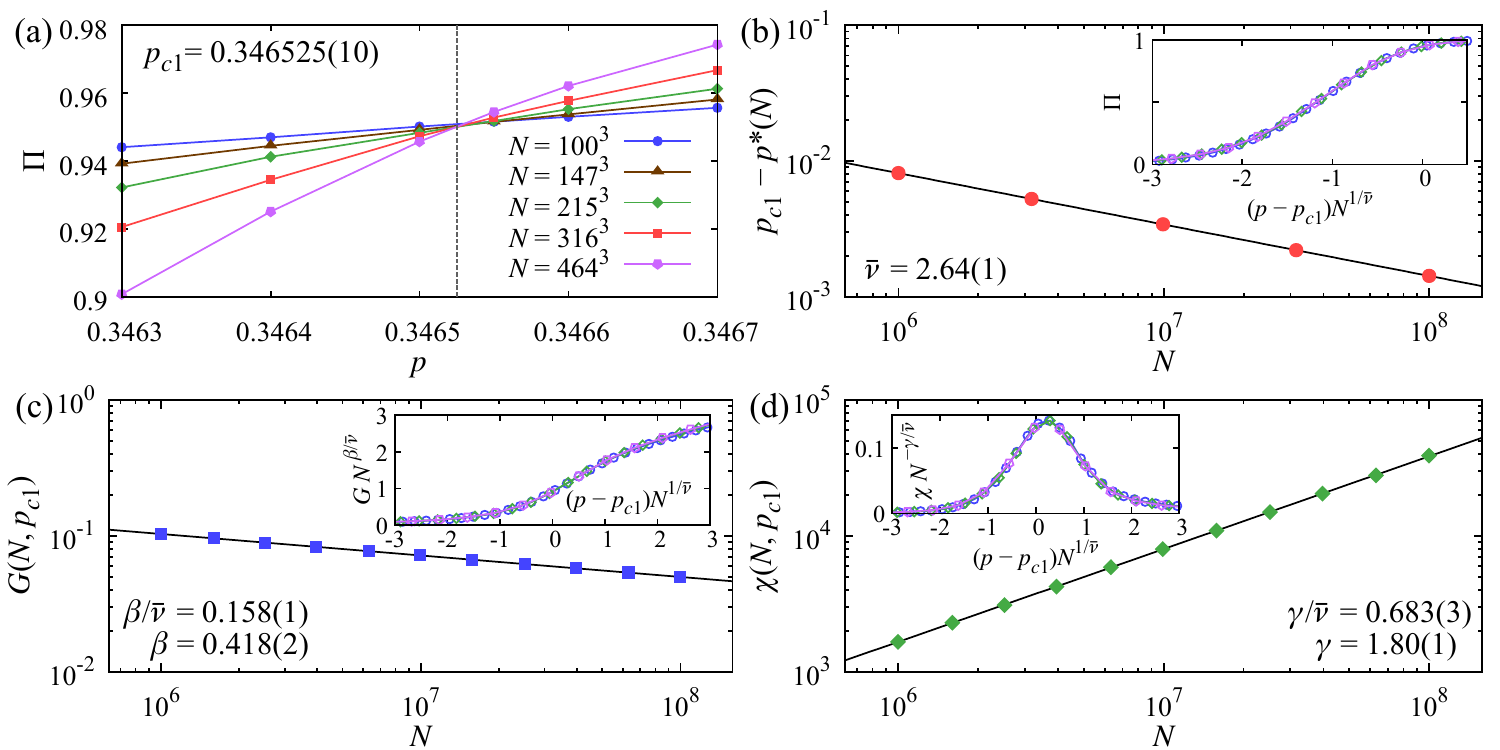}
\vspace*{-0.4cm}
\caption{Finite-size scaling analysis results for 4-selective percolation on the simple cubic lattices at a continuous phase transition point $p_{c1}=0.346\,525(10)$.
(a) Plots of the percolation probability $\Pi$ as a function of $p$.
We used a crossing point of the $\Pi$ for various system sizes as the critical point.
(b) represents the relation of Eq.~(\ref{Eq:FSSCubic41n}), (c) represents the relation of Eq.~(\ref{Eq:FSSCubic41bn}), and (d) represents the relation of Eq.~(\ref{Eq:FSSCubic41gn}).
Points represent Monte Carlo simulation results, and black solid lines are the fitting lines.
(Insets) Plots of finite-size-scaled data collapse curve from $\Pi=\widetilde{\Pi}[(p-p_{c})N^{1/\bar\nu}]$ and Eqs.~(\ref{Eq:CubicScalingAnsatz1}--\ref{Eq:CubicScalingAnsatz2}), using the obtained critical point and critical exponents.
Monte Carlo simulation results with system size $N=100^{3}$ (circle), $215^{3}$ (diamond) and $464^{3}$ (pentagon) are used for the inset plots.}
\label{Fig:FSSCubic01}
\centering
\includegraphics[width=0.95\textwidth]{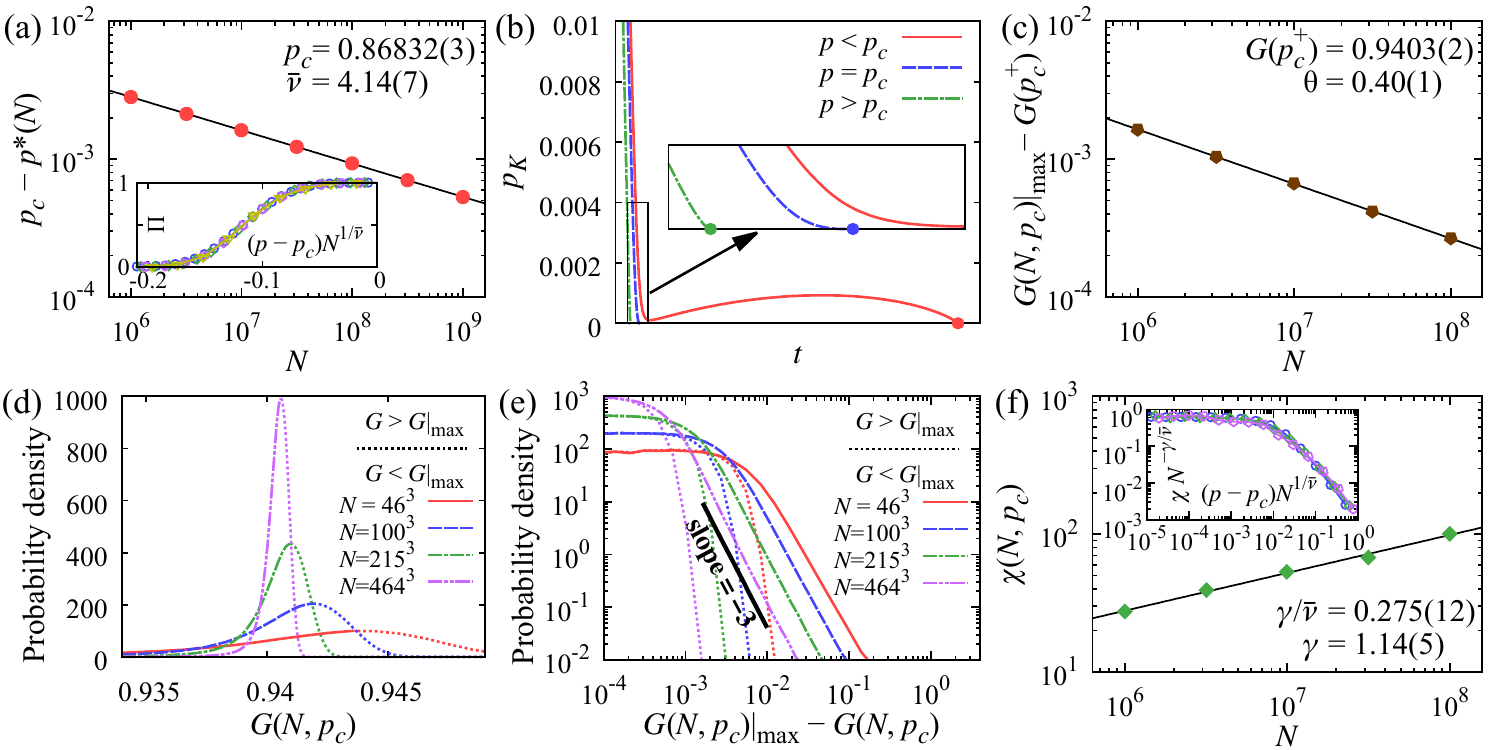}
\vspace*{-0.4cm}
\caption{Finite-size scaling analysis results for 3-selective percolation on the simple cubic lattices at an abnormal discontinuous phase transition point $p_{c}=0.868\,32(5)$ with $G(p^{+}_{c})=0.9403(2)$.
(a) represents the relation of Eq.~(\ref{Eq:FSSCubic31n}).
(b) Plots of $p_{K}$ during the $K$-selective percolation process near the critical point.
Time $t$ has an arbitrary unit.
(c) represents the relation of Eq.~(\ref{Eq:FSSCubic31G}).
(d--e) Plots of the probability density of the order parameter $G$ from the Monte Carlo simulation at the critical point.
(d) uses a normal scale and (e) uses a double logarithm scale with the same data.
(f) represents the relation of Eq.~(\ref{Eq:FSSCubic31gn}).
For (c--f) we only used Monte Carlo ensembles that have the percolating cluster.
(a, c, f) Points represent Monte Carlo simulation results, and black solid lines are the fitting lines.
(Insets) Plots of finite-size-scaled data collapse curve from $\Pi=\widetilde{\Pi}[(p-p_{c})N^{1/\bar\nu}]$ and Eq.~(\ref{Eq:CubicScalingAnsatz2}), using the obtained critical point and critical exponents.
Monte Carlo simulation results with system size $N=100^{3}$ (circle), $215^{3}$ (diamond), $464^{3}$ (pentagon) and $1000^{3}$ (triangle)  are used for the inset plots.}
\label{Fig:FSSCubic02}
\end{figure}
\cleardoublepage
\twocolumngrid
}

\subsubsection{Discontinuous phase transition}
\label{Sec:Discontinuous}
For 3-selective percolation on the simple cubic lattices, we observed discontinuous phase transition into eventual disappearance of the giant component at $p_{c}=0.868\,32(3)$ [Fig.~\ref{Fig:Result02}(b)] with abnormal order parameter distribution and fluctuation.
We obtained the critical point $p_{c}$ and $\bar\nu$ from the following assumption:
\begin{eqnarray}
\label{Eq:FSSCubic31n}
p_{c}-p^{*}(N) &\propto& N^{-1 / \bar\nu}.
\end{eqnarray}
We estimated the critical point $p_{c}=0.868\,32(3)$ and $\bar\nu=4.14(7)$ and the rescaled data using obtained $p_{c}$ and $\bar\nu$ are collapsed well onto a single line [Figs.~\ref{Fig:FSSCubic02}(a)].
The critical exponent $\bar\nu$ obtained from the relation of Eq.~(\ref{Eq:FSSCubic31n}) is consistent with obtained $\bar\nu$ from the relation of $d\Pi(N,p)/dp|_{p=p^{*}} \propto N^{1 / \bar\nu}$, but its value has greater numerical error, therefore we chose the former value.
For $k$-core percolation with $k=4$ on three-dimensional simple cubic lattices with finite $N$, the finite-size behavior of $p_{c}(N)$ is reported to be $p_{c}-p_{c}(N) \sim 1/\log[\log(N^{1/3})]$~\cite{1990EnterFinite, 1999CerfFinite}.
However, this relation does not fit the $K$-selective percolation results.

We concluded this phase transition is discontinuous on the basis of the $p_{3}$ curve [Figs.~\ref{Fig:FSSCubic02}(b)].
For $p>p_{c}$ (red line) it has S-shape, and at $p=p_{c}$ (blue dashed line) it touches zero at the first local minimum.
These results suggest that discontinuity exists in the same way as in ER networks [Fig.~\ref{Fig:Result01}(c)].

Interestingly, the probability density of the order parameter for percolated realizations at the critical point $p_{c}$ has a power-law tail with decay exponent $-3$ for $G(N, p_{c})<G(N, p_{c})|_{\text{max}}$ where $G(N, p_{c})|_{\text{max}}$ is the value of $G(N, p_{c})$ which has maximum probability density [Figs.~\ref{Fig:FSSCubic02}(d--e)].
We also estimated $G(p^{+}_{c})=0.9403(2)$ from the following assumption [Fig.~\ref{Fig:FSSCubic02}(c)]:
\begin{eqnarray}
\label{Eq:FSSCubic31G}
G(N, p_{c})|_{\text{max}} - G(p^{+}_{c}) &\propto& N^{-\theta}.
\end{eqnarray}
Also, the order parameter fluctuation for percolated realization increases as the system size $N$ increases as shown in Fig.~\ref{Fig:FSSCubic02}(f).
These results suggest that order parameter fluctuation diverges in the thermodynamic limit.
Therefore, we tried to obtain other critical exponents from the following relations, which come from the Eqs.~(\ref{Eq:CubicScalingAnsatz1}--\ref{Eq:CubicScalingAnsatz2}):
\begin{eqnarray}
\label{Eq:FSSCubic31bn}
G(N, p_{c})-G(p^{+}_{c}) &\propto& N^{-\beta / \bar\nu},\\
\label{Eq:FSSCubic31gn}
\chi(N, p_{c}) &\propto& N^{\gamma / \bar\nu}.
\end{eqnarray}
The value of $G(N, p_{c})-G(p^{+}_{c})$ in Eq.~(\ref{Eq:FSSCubic31bn}) is comparable to our numerical error of $G(p^{+}_{c})$, thereby critical exponent $\beta$ could not be obtained reliably, nor could we conclusively argue that this phase transition would be hybrid phase transition like on ER networks.
We obtained $\gamma=1.14(5)$ using the relation of Eq.~(\ref{Eq:FSSCubic31gn}) and the rescaled data using obtained $\bar\nu$ and $\gamma$ are collapsed well onto a single line [Figs.~\ref{Fig:FSSCubic02}(f)].
Note that for Figs.~\ref{Fig:FSSCubic02}(c-f), we only use Monte Carlo ensembles that have the percolating cluster.

\section{Conclusion}
\label{Sec:Conclusion}
We studied how the intermediate-degree nodes play a role in the robustness of complex networks, through the $K$-selective percolation model.
We observed the emergence of the fragile valley and resurgent hill in the giant component size $G$, a measure of network robustness.
This unique feature implies that against the selective attack, the complex networks with fewer edges may sometimes be more robust.
We also observed double and even reentrant phase transitions consisting of a series of the hybrid phase transition and the continuous phase transition.
It should be noted that both continuous and hybrid phase transitions were produced by just one simple rule.
A similar model recently studied~\cite{2019SellittoSelective} reported qualitatively different results from ours.
In that study, there appears no valley-hill structure.

In the theoretical perspective, we obtain a new set of critical exponents at a tricritical-like point on ER networks and an abnormal discontinuous transition point on simple cubic lattices.
Particularly, $K$-selective percolation provides another more general schema to induce hybrid and discontinuous phase transitions~\cite{2015D'SouzaAnomalous, 2018LeeRecent, 2019D'SouzaExplosive} than $k$-core percolation.
Moreover, one can observe discontinuous transition at nonzero $q_{c}$ in three-dimensional simple cubic lattices for the 3-selective percolation process.
This is to be compared with the previous results that there is no discontinuous phase transition in cascading failure on multiplex networks with three-dimensional lattices~\cite{2011SonPercolation, 2015GrassbergerPercolation}, or the recent proof that for $k$-core percolation on lattices, discontinuous phase transition appears only at $q_{c}=0$ with $k>d$~\cite{1992SchonmannOn}, where $d$ is a spatial dimension.
This proof for $k$-core percolation however cannot be extended to $K$-selective percolation straightforwardly.
In this regard, $K$-selective percolation could provide a platform for understanding the discontinuous percolation phase transition in low dimensions~\cite{2013ChoAvoiding}.

\section*{Acknowledgement}
This work was supported in part by the National Research Foundation of Korea (NRF) grant funded by the Korea government (MSIT) (No. NRF-2020R1A2C2003669).

\appendix
\section{Polymer network model}
\label{App:Polymer}
In the paper, we used the polymer network model proposed by Kryven \textit{et al.}~\cite{2016KryvenRandom}.
Here we briefly summarize the model description for the sake of self-containedness and reproducibility.
In this model, there are initially $N$ monomers (nodes), and each step polymerization (link formation) occurs.
Each monomer has a fixed number of functional groups (stubs) with probability distribution from real chemical data~\cite{2016KryvenRandom}.

They defined the steric hindrance effect of each node $i$ as $g_{i}$ for the aggregation process between components.
$g_{i}$ has a range from 0 (fully obscure) to 1 (hardly obscure).
If the degree of node $i$ is zero, then $g_{i}=1$.
To find $g_{i}$ for every node with the nonzero degree, matrices $\textbf{A}$, $\textbf{A}'$, and $\textbf{Q}$ are introduced.
$\textbf{A}$ is a adjacency matrix, and the matrices $\textbf{A}'$ and $\textbf{Q}$ are defined as
\begin{equation}
\textbf{A}'=\textbf{A}+\textbf{I},~~~Q_{i, j}=\frac{d_{i}A'_{i, j}}{\sum_{k}(d_{j}A'_{k, j})},
\end{equation}
where $\textbf{I}$ is an identity matrix, and $d_{i}$ is a degree of node $i$.
Finally, they calculated $\textbf{g}^{*}$ by solving the following self-consistence equation, 
\begin{equation}
\textbf{g}^{*} = \textbf{Q} \textbf{g}^{*}.\\
\end{equation}
For multiple components, we need to normalize $g_{i}$ for each component as follows,
\begin{equation}
{g}_{i} = \frac{\text{min} (g_{j}^{*})}{g_{i}^{*}},~~j \in \text{same component with node } i.\\
\end{equation}

They defined the probability of a self-avoiding cyclic chain configuration for cyclization in one component as
\begin{equation}
\Phi(s)_{i, j}=
\begin{cases}
0,~~&s<2.\\
Cs^{-3/2}\exp(-\frac{3}{2}s^{-1} - \alpha s^{1/2}),~~&s\geq2.\\
\end{cases}
\end{equation}
where $C$ is the normalization constant to satisfy $\Phi(2)_{i, j}=1$, $s$ is the length of the shortest path from node $i$ to $j$, and $\alpha$ is one of the model parameters.

Finally, the probability matrix $\textbf{P}$ is defined as follows:
\begin{equation}
P_{i, j}=D \times
\begin{cases}
c_{a} d_{i, \text{free}} d_{j, \text{free}} g_{i} g_{j},&i, j \notin \text{same component.}\\
c_{c} d_{i, \text{free}} d_{j, \text{free}} \Phi(s)_{i, j},&i, j \in \text{same component.}\\
\end{cases}
\label{Eq:P}
\end{equation}
where $D$ is the normalization constant to satisfy $\sum_{i>j} P_{i, j}=1$, $c_{a}$ is aggregation propensity coefficient, $c_{c}$ is cyclization propensity coefficient and $d_{i, \text{free}}$ is the number of remaining stubs of node $i$.
Every step, the probability matrix $\textbf{P}$ is recalculated and a link is made with those probabilities until there remain no more stubs.
We used $N=5\times10^{3}$, $\alpha=4$, $c_{a}=6 \times 10^{-8}$, and $c_{c}=1 \times 10^{-4}$ for the results in the paper.

\bibliography{K-selective_percolation}
\end{document}